\documentclass{article}
\usepackage{color}
\usepackage{amssymb}
\usepackage{amsmath}
\usepackage{latexsym}
\usepackage{slashed, amsthm}

\title{Hidden symmetries and Killing spinor of 5-dimensional minimal gauged supergravity solutions}
\author{C. ~Rugina$^{ \ 1,2}$
	\\
	\\ \small{1. Department of Theoretical Physics, IFIN-HH, Magurele, Romania,}
	\\ \small{2. Department of Physics, University of Bucharest, Bucharest, Romania}
	\\}

\begin{document}
	
	\maketitle 
	{\let\thefootnote\relax\footnotetext{{\em Emails}:
			christina.rugina11@alumni.imperial.ac.uk}}

\begin{abstract}
We investigate the hidden symmetries and existence of Killing spinors of D=5 minimal gauged supergravity solutions which admit a Killing-Maxwell system in the sense of Carter, such as the Chong-Cveti$\check{c}$-L\"{u}-Pope (CCLP) Kerr-dS black hole spacetime and a 5-dimensional minimal gauged supergravity solution endowed with a Sasaki structure deformed by torsion, a limiting case of the generalization of the 5-dimensional toric Sasaki-Einstein $L^{a,b,c}$ spacetime. We note that when an electromagnetic tensor is present and an associated Killing-Maxwell system can be constructed in the sense of Carter, the Killing-Maxwell field becomes a PCKY (principal conformal Killing-Yano) tensor or a PGCKY (principal generalized conformal Killing-Yano) tensor, the latter in the presence of torsion. We find some new hidden symmetries of the Chong-Cveti$\check{c}$-L\"{u}-Pope black hole, i.e. we construct two more generalized St\"ackel-Killing tensors and the associated generalized Killing-Yano (GKY) tensors.  We also explicitly construct a Killing spinor for the specific 5 dimensional minimal gauged supergravity solution that is endowed with a Sasaki structure deformed by torsion mentioned above. 
\end{abstract}

\vskip0.7cm

\textbf{Keywords:} 5-dimensional minimal gauged supergravity; hidden symmetries; Killing spinor; St\"{a}ckel-Killing tensor; Kerr-dS black hole; Killing-Yano tensor; Sasaki deformed by torsion
\vskip0.7cm

\section{Introduction}

Symmetries in nature have long been useful in determining constants of motion and hence in helping solve equations of motion. Hidden symmetries, the symmetries of the phase space, together with spacetime symmetries bring insight into the evolution of a system in curved spacetime and are characterized by symmetrical St\"{a}ckel-Killing (SK) and antisymmetrical Killing-Yano (KY) tensors \cite{1}, while the latter symmetries are driven by Killing vectors.

\medskip
\noindent
There is a long history of separation of variables of Hamilton-Jacobi, Klein-Gordon, and Dirac equations \cite{2,3,4,5,6,7,8,9,10,11,12} against various background spacetimes, starting with the Kerr spacetime in four dimensions and continuing with higher dimensional black hole spacetimes, the work cited above having a particular emphasis on the most general Kerr-AdS-NUT spacetime. All these with the miraculous help of SK and KY tensors, which build up constants of motion, which in turn lead to complete integrability of the above-mentioned equations. Another important object in describing hidden and spacetime symmetries is the PCKY and more generally the CKYs (conformal Killing-Yano tensors) \cite{13,14,15,16}. The PCKY  generates in higher dimensional spacetimes towers of KY and SK tensors, playing an important role in unveiling more hidden symmetries in the respective spacetime.

\medskip
\noindent
The group structure and algebras generated by Dirac-type operators in spacetimes that are torsionless have been thoroughly studied for instance in \cite{17,18,19}: these are operators constructed with the use of KY tensors.

\medskip
\noindent
One other important aspect is the fact that there exist geometrical dualities that map spacetimes with torsion in duals that are torsionless \cite{20,21,22}. If a Killing-Yano structure exists in the spacetime with torsion, then it becomes the vielbein of the torsionless dual spacetime, while the vielbein of the spacetime with torsion becomes the Killing-Yano structure of the torsionless dual.

\medskip
\noindent
A lot of progress \cite{23} has been accumulated since the seminal paper of Myers and Perry \cite{24}, where the metrics describing the isolated, vacuum, rotating higher-dimensional Kerr black holes were derived. Going forward in time we reach the derivation of the most general symmetric non-extremal charged, rotating black-hole in D=5 minimal gauged supergravity spacetime metric \cite{25} which is one of the objects of attention in this current paper. We cannot do justice here to the entire scientific effort of determining the metrics of various black holes in higher dimensions since 1986 on.

\medskip
\noindent
Solving the equations of motion in spacetimes of black holes with gauge fields is aided by the so-called GKY, GSK and GCKY tensors corresponding to symmetries with torsion. There is an ample span of study of these tensors in the literature ranging from the D=5 minimal gauged supergravity spacetime \cite{26,27}, to the study of the black hole spacetimes in the framework of string theory \cite{28,29,30}. 

\medskip
\noindent
Here we study the properties of two related solutions of the D=5 minimal gauged supergravity, with its bosonic string lagrangian in five dimensions. Since there was in the past a lot of work on higher dimensional vacuum solutions\cite{2,3,39,40}, where regular SK and KY tensors helped reveal hidden symmetries of vacuum spacetimes, there is now a lot of work done in recent years on solutions of various supergravities with gauge fields and cosmological constant \cite{41,42,43}. We find here two additional rank -2 generalized closed conformal Killing-Yano tensors, which help find the complete tower of hidden symmetries and isometries of the black hole spacetime. It is also well known that these tensors help classify the higher dimensional spacetimes with gauge fields and also help obtain exact solutions of Einstein equations \cite{reviewHouriYasui}.

\medskip
\noindent
In this paper we discuss the hidden symmetries of a solution of supergravity where the *F (the Hodge dual of the Maxwell tensor) is assimilated with torsion. Hence we work with generalized KY and SK tensors, which in turn can now help separate modified by torsion Dirac and Klein-Gordon equations. The particle equations are different in the presence of gauge fields, as expected and there is an entire theory developed for particles with spin in the presence of torsion \cite{22}.

\medskip
\noindent
The CCLP spacetime we study is endowed with a PGCKY and that is the Killing-Maxwell tensor, the Hodge dual of which is the torsion of the spacetime, a GKY tensor. Spacetimes admitting a PGCKY or a PCKY are remarkable, for instance the Kerr-NUT-AdS spacetime is the only Einstein spacetime admitting a PCKY \cite{reviewHouriYasui}.

\medskip
\noindent
The motivation to study the D=5 minimal gauged supergravity stems from the fact that this is the low energy compactification of type IIB superstring theory and moreover its properties match the 4-dimensional boundary of the $AdS_5$ in AdS/CFT correspondence. So our task here is to extend the work done in finding the hidden symmetries of the most general non-extremal rotating black hole solution of the D=5 minimal gauged supergravity theory, the Chong-Cveti$\check{c}$-L\"{u}-Pope black hole spacetime by \cite{26} and find new GSK and GKY tensors for it. Our work is inscribed together with previous work to extend the study of higher dimensional vacuum solutions to those with gauge fields and cosmological constant. The result of this study partially answers to interesting questions related to the significance of GKY and GSK tensors, their relations to generalized hidden symmetries and algebraic type of solutions  \cite{26}. 

\medskip
\noindent
The Kerr-dS black hole in 5-dimensional minimal gauged supergravity, the CCLP black hole is also a Sasaki deformed by torsion metric \cite{50}, so it has that in common with the Houri-Takeuchi-Yasui metric \cite{50} for which we find a Killing spinor and which is a generalization of the 5-dimensional toric Sasaki-Einstein spacetime $L^{a,b,c}$, obtained by basically taking a certain limit of a generalization of the global metric of $L^{a,b,c}$. Sasaki with torsion Euclidean versions of Lorentzian 5-dimensional minimal gauged supergravity can be obtained via a Wick rotation.

\medskip
\noindent
In section \ref{sect2} we start by presenting some relevant calculus relations with torsion. In section \ref{sec3} we introduce some results obtained by Carter in the context of the Kerr-Newman black-hole, dubbed the Killing-Maxwell system, and we also review some results obtained previously in \cite{17} where Dirac-type operators were constructed to describe hidden symmetries (here we write them out specifically in the framework of the Killing-Maxwell system of the Kerr-Newman black hole). We then generalize in section \ref{sec4} Carter's Killing-Maxwell system in the context of D=5 minimal gauged supergravity (new result), we review the structure of the  CCLP black hole and we find two new GSK and GKY tensors for this spacetime. In section \ref{sec5} we explicitly find a Killing spinor for a specific manifold endowed with a Sasaki structure deformed by torsion, that is a solution of minimal gauged supergravity in 5 dimensions. In conclusion, we introduce in this paper two new results with direct application in 5-dimensional minimal gauged supergravity. 

\section{Some calculus relations in the presence of torsion}
\label{sect2}

In the following, the KY tensor $Y \in \Omega^p(\mathcal{M})$, the 
SK tensor $K$, and the CKY tensor (also denoted $Y$) on a pseudo-Riemannian manifold $(\mathcal{M},g)$ are used with their usual  definitions (for example see \cite{31}). These definitions also hold in the case of a manifold with torsion. In this case  the tensors are called, as usually, generalized, and one just has to write the meaningful connection for this case. 

\medskip

\noindent
The basic calculus formulas in presence of torsion $T$ can be found, for example, in \cite{31,32,33,34,35}). Let T be a 3-form on a pseudo-Riemannian manifold $(\mathcal{M},g)$ and ${e_a}$ an orthonormal frame such that $g(e_a, e_b)=\delta_{ab}$. Then if X, Y are vector fields then we define the Levi-Civita connection as:

\begin{equation}
\nabla^T_X Y = \nabla_X Y +\frac{1}{2} T(X,Y,e_a) e_a,
\end{equation}

\noindent
T is assimilated with torsion and we shall also use the T=2A notation for torsion. The covariant derivative for  p-form $\omega$ is:

\begin{equation}
\nabla^A_X \omega = \nabla_X \omega - (X \lrcorner e_b \lrcorner A) \wedge (e_b \lrcorner \omega),
\end{equation}

\noindent
with the explicit formula for a 2-form being:

\begin{equation}
\nabla^A_\mu Y_{\nu\rho} = \nabla_\mu Y_{\nu\rho} - 2 A_{\sigma\mu[\nu} {Y^{|\sigma|}}_{\rho]}.
\end{equation}

\noindent
Note that the spinor covariant derivative with torsion can be written out as:

\begin{equation}
D^A_\mu = D_\mu + \frac{1}{12} \gamma^\nu \gamma^\rho A_{\mu\nu\rho}.
\end{equation}

\noindent
Hence the Dirac operator with torsion is:

\begin{equation}
D^A_\mu \gamma^\mu = D_\mu \gamma^\mu + \frac{1}{12} \gamma^\mu \gamma^\nu \gamma^\rho A_{\mu\nu\rho}.
\end{equation}

\noindent
The Ricci relation with torsion for a Killing-Yano 2-form is as follows:

\begin{equation}
\nabla^A_\alpha \nabla^A_\beta Y_{\mu\nu} = -\frac{3}{2}{R^\lambda}_{\alpha\beta[\mu} Y_{\nu]\lambda} - 2 {A^{|\lambda|}}_{\beta[\alpha} \nabla^A_{|\lambda|} Y_{\mu]\nu}.
\end{equation}

\noindent
The square of the Dirac operator as a function of the torsion T is:

\begin{equation}
{D^2}^T = -\Delta^T -\frac{dT}{4} -\frac{s}{4} -\frac{||T||^2}{24},
\end{equation}

\noindent
where

\begin{equation}
\Delta^T = \nabla^T_{X_a} \nabla^T_{X^a} + \nabla^T_{\nabla^T_{X_a} X^a},
\end{equation}

\noindent and s is the scalar curvature of the connection with torsion:

\begin{equation}
s = - X^a \lrcorner R(X_a, X_b) e^b.
\end{equation}

\noindent 
The curvature operator is defined as usual:

\begin{equation}
R(X,Y)\omega = (\nabla^T_X \nabla^T_Y - \nabla^T_Y \nabla^T_X -\nabla^T_{[X,Y]}) \omega.
\end{equation}

\noindent
So the commutator of the spinor covariant derivatives is:
\begin{equation}\label{anticommDirac}
[D^A_\mu, D^A_\nu] \Psi = \frac{1}{8} R_{\alpha\beta\mu\nu} [ \gamma^\alpha, \gamma^\beta] \Psi - {A^\lambda}_{\mu\nu} D^A_\lambda \Psi.
\end{equation}

\section{Review of Carter's 4-dimensional Killing-Maxwell System}
\label{sec3}

The minimal gauged supergravity in 5 dimensions spacetime together with its symmetries has been recently studied in \cite{26} and in this framework- also the most general known spherical symmetric charged rotating black hole solution, the Chong-Cveti$\check{c}$-L\"{u}-Pope black hole \cite{25}. Previous work in 4 dimensions dates back to 1987 and was done by Carter \cite{36,37} who investigated the solutions and symmetries of its lower-dimensional cousin, the Kerr-Newman black hole. Carter reached the conclusion that there exists a Killing-Maxwell electromagnetic system defined by the following equation for the 4-dimensional electromagnetic potential (here we used Carter's notations, in that semi-colon means taking the covariant derivative):

\begin{equation}
\hat{A}_{[\mu;\nu];\rho} = 2 \frac{4\pi}{3}\hat{j}_{[\mu}g_{\nu]\rho}, \label{K-Mpotential}
\end{equation}

\noindent
where g is the spacetime metric and $\hat{j}$ the current, which is a primary Killing vector (for definition, see for instance \cite{13}). The Killing-Maxwell electromagnetic system obeys regular Maxwell equations, since the Maxwell field is defined as usual and respects Maxwell's laws:

\begin{equation}
\hat{F}^{\rho\mu}_{;\rho} = 4\pi\hat{j}^\mu \label{PCKY}
\end{equation}

\noindent
and

\begin{equation}
\hat{F}_{[\mu\nu;\rho]} = 0. \label{Max}
\end{equation}

\noindent
Further, Carter mentions that the Hodge dual of the Killing-Maxwell electromagnetic field is a Killing-Yano tensor of rank 2. Although he doesn't state it directly, it follows that $\hat{F}$ is a PCKY tensor (principal conformal Killing-Yano tensor)- as defined for instance by Kubiz$\check{n}\acute{a}$k in his thesis \cite{13}, rel. (3.7)- since it satisfies equations (\ref{K-Mpotential}) and (\ref{PCKY}). The fact that $\hat{F}$ is closed follows from Maxwell's laws (\ref{Max}). 

\medskip
\noindent
The Hodge dual of the Killing-Maxwell electromagnetic field consequently determines a Dirac-type operator, that anti-commutes with the Dirac operator in the Kerr-Newman spacetime, according to \cite{17}:

\begin{equation} \label{Q*F}
Q_{*\hat{F}} = \gamma^\mu {*\hat{F}_\mu}^\nu D_\nu - \frac{1}{6} \gamma^\mu \gamma^\nu \gamma^\rho \nabla_\mu *\hat{F}_{\nu\rho}.
\end{equation}

\noindent
This operator corresponds to a quantum (non-anomalous) hidden symmetry for the spinning point particle in Kerr-Newman spacetime which is - as stated in \cite{17} - an additional non-generic supersymmetry of the particle. It is a remarkable fact that $*\hat{F}$ generates a supersymmetry and it points out the subtle connection between the symmetries of the Killing-Maxwell electromagnetic field in curved spacetime and this supersymmetry and further- between spin and electric charge. In the 4-dimensional spacetime, the PCKY generates one Killing-Yano tensor, which is $*\hat{F}$ (the Hodge dual of a closed conformal Killing-Yano tensor is a Killing-Yano tensor) and one St\"{a}ckel-Killing tensor, K:

\begin{equation}
K_{\mu\nu} = (*\hat{F})_{\mu\rho} {(*\hat{F})}_\nu^\rho.
\end{equation}

\section{Hidden symmetries of the D=5 minimal gauged supergravity spacetime with a Killing-Maxwell system}
\label{sec4}

In the cousin spacetime, D=5 minimal gauged supergravity, the Hodge dual of the electromagnetic field plays the role of torsion as evidenced in \cite{26}. Here we are going to focus on the case when *F is part of the Killing-Maxwell system, a generalization to 5 dimensions of the work set forth by Carter. This means we are going to identify the PGCKY of the 5-dimensional spacetime, of which *F is a Killing-Yano tensor, with F (the Killing-Maxwell electromagnetic field). The definition of a Killing-Maxwell electromagnetic field in D=5 supergravity with torsion is (index A stands for torsion):

\begin{equation}
\nabla_\rho^A F_{\mu\nu} = 2 \pi  g_{\rho[\mu} j_{\nu]} \label{eq_long_KMax}
\end{equation} 

\noindent
and together with:

\begin{equation}
\nabla_\rho^A F^{\rho\mu} = 4\pi j^\mu 
\end{equation}

\noindent
form a Killing-Maxwell system that obeys Einstein-Maxwell's laws. According to \cite{26} the PGCKY of the D=5 minimal gauged supergravity spacetime is defined as:

\begin{equation}
\nabla_\rho h_{\mu\nu} = 2g_{\rho [\mu} \xi_{\nu]} -\frac{1}{\sqrt{3}} (*F)_{\rho \sigma[\mu}{h^\sigma }_{\nu]}. \label{eq_KMax}
\end{equation} 

\noindent
So the equation~(\ref{eq_KMax}) becomes:

\begin{equation}
{\nabla^A}_\rho h_{\mu\nu} = 2g_{\rho [\mu} \xi_{\nu]}.
\end{equation} 

\noindent
Note that if in the equation above we notate h by F and we set $j = \pi \xi$ then indeed the definition in 5-dimensions of a Killing-Maxwell electromagnetic field and that of a PGCKY coincide and hence our assumption that for the five-dimensional supergravity endowed with a Killing-Maxwell system, such that $*F$ is Killing-Yano, then  F and h coincide is true. According to the theory of Killing-Yano tensors, $* F $ the Killing-Yano tensor is the Hodge dual of the Killing-Maxwell electromagnetic field and which also plays the role of torsion in this spacetime. It also generates the St\"{a}ckel-Killing tensor:

\begin{equation}
K_{\mu\nu} = (*F)_{\mu \rho \sigma} (*F)_\nu^{\rho \sigma}.
\end{equation}

\medskip
\noindent
Let's now take a look at the Chong-Cveti$\check{c}$-L\"{u}-Pope black hole (CCLP) in D=5 minimal gauged supergravity framework, with the following notations:

\begin{equation}
g = \sum_{\mu =x,y} (\omega^\mu \omega^\mu + \tilde{\omega}^\mu \tilde{\omega}^\mu) + \omega^\epsilon \omega^\epsilon,
\end{equation}

\begin{equation}
A = \sqrt{3} (A_q + A_p).
\end{equation}

\noindent 
And-

\begin{equation}
\omega^x = \sqrt \frac{x-y}{4X} dx, \hspace{0.5in} \tilde{\omega^x} = \frac{\sqrt{X} (dt + y d\phi)}{\sqrt{x(y-x)}},
\end{equation}

\begin{equation}
\omega^y = \sqrt{\frac{y-x}{4Y}} dy, \hspace{0.5in} \tilde{\omega^y} = \frac{\sqrt{Y} (dt + x d\phi)}{\sqrt{y(x-y)}},
\end{equation}

\begin{equation}
\omega^\epsilon = \frac{1}{\sqrt{-x y}}[\mu dt +\mu(x+y) d\phi +xy d\psi -y A_q
-xA_p],
\end{equation}

\begin{equation}
A_q = \frac{q}{x-y} (dt + y d\phi), \hspace{0.5in} A_p =\frac{-p}{x-y} (dt + x d\phi),
\end{equation}

\noindent
and

\begin{equation}
X = (\mu + q)^2 + Ax + CX^2 + \frac{1}{12}\Lambda x^3,
\end{equation}

\begin{equation}
Y = (\mu + p)^2 + By + Cy^2 + \frac{1}{12} \Lambda y^3.
\end{equation}

\noindent
Please note that it was proven in \cite{30} that CCLP is the unique minimal gauged supergravity spacetime with torsion such that the torsion tensor is both closed ($d^T T=0$ ) and co-closed ($\delta^T T=0$).
We then find following \cite{26} that, F, the PGCKY in our case is:

\begin{equation}\label{eq:F}
F = \sqrt{-x} \tilde{\omega}^x \wedge \omega^x +\sqrt{-y} \tilde{\omega}^y \wedge \omega^y
\end{equation}

\noindent
and the corresponding Killing tensor:

\begin{equation}
K = y(\omega^x \omega^x +\tilde{\omega}^x \tilde{\omega}^x) + x(\omega^y \omega^y +\tilde{\omega}^y \tilde{\omega}^y) +(x+y)\omega^\epsilon \omega^\epsilon.
\end{equation}

\noindent
Note that K is directly involved in separating the Hamilton-Jacobi and Klein-Gordon equations in this spacetime.

\medskip
\noindent
We can now proceed to write down a couple of new GCCKY (generalized closed conformal Killing-Yano) tensors for the above metric:

\begin{equation}
h_1 = \sqrt{4X} \tilde{\omega^x} \wedge \omega^\epsilon + \sqrt{Y} \tilde{\omega^y} \wedge \omega^\epsilon
\end{equation}

\noindent
In a similar way we can retrieve:

\begin{equation}
h_2 = \sqrt{-xy} \omega^x \wedge \omega^\epsilon + \sqrt{-xy} \omega^y \wedge \omega^\epsilon
\end{equation}

\noindent
$\omega$ is a Darboux basis for $h_1, h_2$ and h and so that means that h, $h_1, h_2$ are linear independent and so the ancillary GSK and GKY tensors are linearly independent, too.

The tensors GCCKY above generate 2 new St\"{a}ckel-Killing tensors via the following relation:

\begin{equation}
K_{ab} = h_{ac} {h_b}^c -\frac{1}{2} g_{ab} h^2
\end{equation} 

\noindent
which are

\begin{equation}\label{eq:stackel_psi}
K_1 = (x-y) \tilde{\omega^y}\omega^\epsilon + x \tilde{\omega^x}\omega^y,
\end{equation}

\noindent
respectively,

\begin{equation}\label{eq:stackel_phi}
K_2 =  (x-y) \tilde{\omega^x}\omega^\epsilon + y \omega^x \tilde{\omega^y}.
\end{equation}

\medskip
\noindent
Also, for spacetimes with $D \ge 6$ where an electromagnetic field is present and a Killing-Maxwell system can be constructed, the PGCKY is, naturally, in these spacetimes still the Killing-Maxwell electromagnetic tensor, F, which generates a tower of Killing-Yano and St\"{a}ckel-Killing tensors as follows:

\begin{equation}
F^{(j)} = F \wedge \cdots \wedge F.
\end{equation}

\noindent
Above the wedge is taken j times, $F^{(j)}$ is a (2j)-form and $F^{(1)} = F$. $F^{(j)}$ is a set of (n-1) non-vanishing closed CKY (conformal Killing-Yano) tensors, where D, the dimension of the spacetime is-

\begin{equation}
D = 2n +\epsilon.
\end{equation}

\noindent
Here $\epsilon =0,1$ depending on whether D is even or odd. And this generates the towers of n-1 rank (D-2j) Killing-Yano tensors:

\begin{equation}
Y^{(j)} = *F^{(j)}
\end{equation}

\noindent
and n-1 rank-2 St\"{a}ckel-Killing tensors:

\begin{equation}
K^{(j)}_{\mu\nu} = Y^{(j)}_{\mu \rho_1 \cdots \rho_{D-2j-1}} {Y^{(j)}_\nu}^{\rho_1 \cdots \rho_{D-2j-1}}.
\end{equation}

\medskip
\noindent
Now let's turn our attention to the spinning point particle in the presence of torsion. The rank-2 quantum phase space Dirac-type operator in the presence of torsion is given in \cite{27}. Here, however,  we are going to focus on writing the associated GKY with the GSK tensors given by $(\ref{eq:stackel_psi}), (\ref{eq:stackel_phi})$ , which we are going to notate with $Y_1, Y_2$. So the two GKY tensors are:

\begin{equation}\label{eq:KY2}
Y_1 = \sqrt{x-y} \omega^x \wedge \omega^y \wedge \tilde{\omega^y} +\sqrt{-x} \omega^x \wedge \tilde{\omega^x} \wedge \omega^y
\end{equation}

and

\begin{equation}\label{eq:KY3}
Y_2 = \sqrt{x-y} \tilde{\omega^x} \wedge \omega^y \wedge \tilde{\omega^y} + \sqrt{-y} \omega^x \wedge \tilde{\omega^y} \wedge \tilde{\omega^x}.
\end{equation}

\noindent
and this concludes our new result for CCLP here.

\section{Killing spinors in D=5 minimal gauged supergravity}
\label{sec5}

\noindent
It is well-known that there is a close intertwining between the existence of Killing spinors and Killing-Yano tensors and other structures and that Killing spinors have been widely used to classify solutions of supergravity in various dimensions. The supersymmetric solutions of minimal gauged and ungauged supergravity were classified for instance in \cite{46,47}. In particular for D=5 the Killing spinors obey the equation:

\begin{equation}\label{eq: KillingSpinorEq}
[D_{\alpha} + \frac{1}{4\sqrt{3}}(\gamma_{\alpha}^{\beta \gamma} -4 \delta_{\alpha}^{\beta} \gamma^\gamma) F_{\beta \gamma}]\epsilon^a - \chi \epsilon^{ab} (\frac{1}{4\sqrt{3}}\gamma_\alpha -\frac{1}{2} A_\alpha)\epsilon^b = 0,
\end{equation}

\noindent
where $\chi$ is a real constant and $\epsilon^{ab}$ the Levi-Civita tensor.

\medskip
\noindent
We are trying to go deeper here and explore the relationship between Killing spinors and Killing-Yano tensors. In that, we focus on a spacetime in 5 dimensions with deformed Sasaki structure by torsion in the context of minimal gauged supergravity described by the Houri-Takeuchi-Yasui metric.

\medskip
\noindent
We can now turn to solutions of eq. (\ref{eq: KillingSpinorEq}) for the case of a manifold with Sasaki deformed by torsion  structure in 5-dimensional minimal gauged supergravity. In D=5 minimal gauged supergravity there exists always a symplectic Majorana spinor \cite{48} and we are going to construct a solution to equation (\ref{eq: KillingSpinorEq}) on the model of the solution given in \cite{49}, where Killing spinors of the following form were constructed in $AdS_5$:

\begin{equation}
\epsilon_i = (e^{\frac{i}{2} a r M \gamma_r})_{ij} (\delta_{jk} + \frac{i}{2} a x^{\alpha} \gamma_{\alpha} (M_{jk} - i \delta_{jk} \gamma_r)) \xi_k.
\end{equation}

\noindent
With this starting point, we formulated our own ansatz that the Killing spinor that verifies equation (\ref{eq: KillingSpinorEq}) is of the form:

\begin{equation}\label{KillingSpinor}
\epsilon_i = {(e^{\frac{i}{2}\gamma^i x_i M})_j}^k ( \delta_i^j x^\alpha({\gamma_\alpha}^{\beta\delta} - \delta_\alpha^\beta \gamma^\delta) F_{\beta \delta} + \frac{i \epsilon^{jl}}{2} \chi x^\alpha \gamma_\alpha (M_{il} - i \delta_{il}A_{\alpha}\gamma^{\alpha}))\xi_k,
\end{equation}

\noindent
where $M= \vec{x} \vec{\sigma}$ with $\vec{\sigma}$ the Pauli matrices and

\begin{equation}
\vec{x} = (\sin \theta \cos \phi, \sin \theta \sin \phi, \cos \theta)
\end{equation}

\noindent
and  $\epsilon^{ij}$ is the Levi-Civita tensors and $\xi_k$ is a symplectic Majorana spinor. We also note that the vector $x_\alpha$  (in the coordinates in which our Sasaki deformed by torsion minimal gauged supergravity metric is given) is given by:

\begin{equation}
x_0 = \psi_0, 
\end{equation}

\begin{equation}
x_1 = x \cos \theta, 
\end{equation}

\begin{equation}
x_2 = x \sin \theta \cos \phi, 
\end{equation}

\begin{equation}
x_3 = x \sin \theta \sin \phi \cos \psi_1, 
\end{equation}

\begin{equation}
x_4 = x  \sin \theta \sin \phi \sin \psi_1.
\end{equation}

\noindent
Moreover $\epsilon_i$ needs to verify rel (5.1) in \cite{46}, the integrability conditions of the Killing spinor and consequently there will be constraints for the Majorana spinors as well, if we plug in the expression for $\epsilon_i$ in rel. (5.1) of \cite{46}. So the integrability condition reads:

\begin{multline}
\{\frac{1}{8} {^5 R}_{\rho \mu \nu_1 \nu_2} \gamma^{\nu_1 \nu_2} +\frac{1}{4 \sqrt 3}({\gamma_{[\mu}}^{\nu_1 \nu_2} + 4 \gamma^{\nu_1} \delta^{\nu_2}_{[\mu}) \nabla_{\rho]} F_{\nu_1 \nu_2} - \\  \\-
\frac{1}{48}(-2F^2 \gamma_{\mu \rho} - 8 F^2_{\nu[\rho} {\gamma^{\nu}}_{\mu]} + 12 F_{\mu \nu_1} F_{\rho \nu_2} \gamma^{\nu_1 \nu_2} + 8 F_{\nu_1 \nu_2} F_{\nu_3[\rho} {\gamma_{\mu]}}^{\nu_1 \nu_2 \nu_3}) - \\ \\ -
\frac{\chi^2}{48} \gamma_{\rho \mu}\} \epsilon^a - \frac{\chi}{24} ({\gamma_{\rho \mu}}^{\nu_1 \nu_2} F_{\nu_1 \nu_2} -4 F_{\nu[\rho} {\gamma_{\mu]}}^\nu - F_{\rho \mu}) \epsilon^{ab} \epsilon^b = 0.
\end{multline}

\noindent
We work with the metric in \cite{50}, which is Sasaki with torsion and satisfies the equations of motion of 5-dimensional minimal gauged supergravity:

\begin{multline}
g = (\xi - x)(d\theta^2 + \sin^2 \theta d \phi^2) +\frac{dx^2}{Q(x)} + Q(x)(d \psi_1 + \cos \theta d\phi)^2 + \\ \\ +
4 (d \psi_0 + (x+ \frac{q}{x-\xi}) d\psi_1 + (x - \xi + \frac{q}{x-\xi}) 
\cos \theta d \phi)^2,
\end{multline}

\noindent
where

\begin{equation}
Q(x) = \frac{4 x^3 + (1-12 \xi) x^2 + (8q -2 \xi + 12 \xi^2) x + k}{\xi- x},
\end{equation}

\noindent
and $q$, $\xi$ and $k$ are free parameters. As it is well known the action in minimal gauged supergravity in 5 dimensions is:

\begin{equation}
\mathcal{L}_5 = *(\mathcal{R} - \Lambda) - \frac{1}{2} F \wedge *F +\frac{1}{3 \sqrt{3}} F \wedge F \wedge A,
\end{equation} 

\noindent
where $F= dA$ and in our case the Maxwell potential is:

\begin{equation}
A = - \frac{2\sqrt{3} q}{x- \xi} (d \psi_1 + \cos \theta d \phi),
\end{equation}

\noindent
and the torsion $T = *F/ \sqrt{3}$. The equations of motion are:

\begin{equation}
R_{ab} = -4 g_{ab} + \frac{1}{2}( F_{ac}{F_b}^c -\frac{1}{6} g_{ab} F_{cd} F^{cd}),
\end{equation}

\begin{equation}
d*F -\frac{1}{\sqrt{3}} F \wedge F = 0.
\end{equation}

\noindent
Consequently, we can derive these results for this metric:

\begin{equation}
F = \frac{2 \sqrt{3} q}{(x-\xi)^2} dx \wedge d\psi_1 + \frac{2\sqrt{3} q}{x-\xi} \sin \theta d\theta \wedge d \phi + \frac{2\sqrt{3}q}{(x-\xi)^2} \cos \theta dx \wedge d\phi.
\end{equation}
\noindent
We defer the calculations to the appendix, where we spell out the inverse metric, the Christoffel symbols, the Ricci tensor and the spin connections. After highly non-trivial work and cancellations, plugging in formula ($\ref{KillingSpinor}$) in eq. $(\ref{eq: KillingSpinorEq})$ gives the much sought-after null result, and this concludes our proof here. One can extend the study of the relationship between the existence of a (generalized) Killing-Yano tensor and that of a Killing spinor to other supergravities in various dimensions, but we leave this for future work.

\section{Conclusions}

\noindent
We overviewed briefly some of the symmetries of the Kerr-Newman spacetime, endowed with a Killing-Maxwell system and higher dimensional minimal gauged supergravity solutions with torsion (such as the CCLP black hole spacetime). For the studied 5-dimensional spacetimes the Hodge dual of the Killing-Maxwell electromagnetic field plays the role of torsion and is at the same time a generalized Killing-Yano tensor of the spacetime, being derived naturally from the PGCKY. We find two more GKY and GSK tensors (generalized Killing-Yano and Stackel-Killing) for the CCLP spacetime and we write out the Dirac operator with torsion. We then turn to another related example of minimal gauged supergravity in 5 dimensions with Sasaki structure deformed by torsion, the Houri-Takeuchi-Yasui metric, and we find a Killing spinor for it.

\medskip
\noindent
It is interesting that indeed supersymmetries are generated by the Killing-Maxwell electromagnetic field when this is present and this sparks further investigation of the correlation between the electromagnetic gauge symmetry and supersymmetry in curved spacetimes, which was pursued initially by Carter. Also an interesting future track would be to determine the dual (torsionless) spacetime of the D=5 minimal gauged supergravity solutions used here.

\vskip1cm
\section*{Acknowledgements}
C.R. acknowledges helpful discussions with Virgil Baran and Andrei Ludu and the support of the POS-DRU European fellowship, which were instrumental in carrying out the current work. The author also likes to thank Gary Gibbons, Tsuyoshi Houri and Aurelian Isar for reading the manuscript at various stages and for making useful comments.

\section{Appendix}

\noindent
This appendix presents calculations for section 5.

\noindent
The inverse metric of the Sasaki with torsion on in section 5 has the following non-zero components:

\begin{equation}
g^{\psi_0 \psi_0} = \frac{(\xi-x)}{D(\psi_0, \psi_1, x, \theta, \phi)} [(\xi - x) sin^2 \theta - 4 Q(x)],
\end{equation}

\begin{equation}
g^{\psi_1 \psi_1} = \frac{4 (\xi-x)^2 sin^2 \theta}{Q(x) D(\psi_0, \psi_1, x, \theta, \phi)},
\end{equation}

\begin{multline}
g^{xx} = \frac{(\xi-x)}{D(\psi_0, \psi_1, x, \theta, \phi)}\{-16 Q^2 (x) cos^2 \theta + 4 Q(x) [(\xi-x) sin^2 \theta + \\ \\ +
32 cos \theta (x +\frac{q}{x-\xi})(x-\xi +\frac{q}{x-\xi}) - 16 (x -\xi +\frac{q}{x -\xi})^2] - \\ \\ -
320 (x +\frac{q}{x-\xi})^2 (x-\xi +\frac{q}{x - \xi})^2  - 48 ( x+ \frac{q}{x -\xi})^2 (\xi- x) sin^2 \theta\},
\end{multline}

\begin{multline}
g^{\theta \theta} = \frac{1}{Q(x) D(\psi_0, \psi_1, x, \theta, \phi)}\{- 8 Q^2(x) cos ^2 \theta + Q(x)[4(\xi -x) sin^2 \theta + \\ \\ +128 cos \theta (x-\xi+\frac{q}{x - \xi}) ( x+ \frac{q}{x- \xi}) - 64 (x -\xi +\frac{q}{x - \xi})^2] - \\ \\-
48 (x+\frac{q}{x-\xi})^2 (\xi-x) sin^2 \theta + 256 (x +\frac{q}{x-\xi})^2 (x-\xi +\frac{q}{x-\xi})^2 \},
\end{multline}

\begin{equation}
g^{\phi \phi} = \frac{4(\xi-x)}{D(\psi_0, \psi_1, x, \theta, \phi) Q(x)} \{Q(x) - 48 (x+\frac{q}{x-\xi})^2\},
\end{equation}

\begin{multline}
g^{\psi_0 \psi_1}= g^{\psi_1 \psi_0} = \frac{(\xi -x)}{Q(x)  D(\psi_0, \psi_1, x, \theta, \phi)} \{ -16 Q(x) cos \theta (x-\xi +\frac{q}{x-\xi}) + \\ \\+ (x+\frac{q}{x-\xi}) (\xi -x) sin^2 \theta - 64 (x + \frac{q}{x-\xi})(x-\xi +\frac{q}{x-\xi})^2 \},
\end{multline}

\begin{equation}
g^{\psi_0 x } = g^{x \psi_0} = 0, \hspace{1.0in} g^{\psi_0 \theta} = g^{\theta \psi_0} = 0,
\end{equation}

\begin{multline}
g^{\psi_0 \phi} = g^{\phi \psi_0} = \frac{x-\xi}{Q(x)D(\psi_0, \psi_1, x, \theta, \phi)} \{-16 Q(x) cos \theta (x -\xi+\frac{q}{x -\xi}) - \\ \\ -
64 ( x- \xi + \frac{q}{x - \xi})^2 (x+\frac{q}{x -\xi}) + 8 (x+\frac{q}{x- \xi})(\xi -x) sin^2\theta\},
\end{multline}

\begin{equation}
g^{\psi_1 x} = g^{x \psi_1} = 0, \hspace{1.0in} g^{\psi_1 \theta} = g^{\theta \psi_1} = 0,
\end{equation}

\begin{multline}
g^{\psi_1 \phi} = g^{\phi \psi_1} = \frac{x-\xi}{Q(x)D(\psi_0, \psi_1, x, \theta, \phi)}\{8 Q(x) cos \theta \\ 
- 32(x+\frac{q}{x- \xi})(x-\xi +\frac{q}{x-\xi}) \},
\end{multline}

\begin{equation}
g^{x \theta} = g^{\theta x} = 0, \hspace{0.5in} g^{x \phi} = g^{\phi x} = 0, \hspace{0.5in}  g^{\theta \phi}= g^{\phi \theta}=0,
\end{equation}

\noindent 
where the function D (the determinant of g) is:

\begin{multline}
D(\psi_0, \psi_1, x, \theta, \phi) =-16 Q^2 (x) cos^2 \theta + Q(x) [4(\xi-x) sin^2 \theta + \\ \\ +
224 (x+ \frac{q}{x-\xi})(x-\xi +\frac{q}{x-\xi}) cos \theta  - 64 (x -\xi  +  \frac{q}{x-\xi})^2] + \\ \\ +
16 (x +\frac{q}{ x-\xi})^2 [-3 (\xi - x) sin^2 \theta  + 32 (x - \xi +\frac{q}{x-\xi})].
\end{multline}

\noindent 
And the torsion tensor is:

\begin{multline}
*F = g^{xx} g^{\psi_1 \psi_1} \frac{2 \sqrt{3} q}{(x-\xi)^2} d\psi_0 \wedge d\theta \wedge d\phi + 
g^{\theta \theta} g^{\phi \phi} \frac{2 \sqrt{3} q}{x-\xi} sin \theta d\psi_0 \wedge d\psi_1 \wedge dx + \\ \\ +
g^{xx} g^{\phi \phi} \frac{2 \sqrt{3} q}{(x-\xi)^2} cos \theta d\psi_0 \wedge d\psi_1 \wedge d \theta .
\end{multline}

\noindent
The inverse vielbeins are:

\begin{equation}
\hat{e}_1 = - g^{\theta \theta} \sqrt{\xi -x} d\theta,
\end{equation}

\begin{multline}
\hat{e}_2 = g^{\phi \phi} \sqrt{\xi-x} sin \theta d \phi + g^{\psi_1 \phi}\sqrt{\xi - x} sin \theta d \psi_1 + g^{\psi_0 \phi} \sqrt{\xi -x} sin \theta d \psi_0,
\end{multline}

\begin{equation}
\hat{e}_3 = g^{xx} \frac{1}{\sqrt{Q(x)}} dx,
\end{equation}

\begin{multline}
\hat{e}_4 = ( g^{\psi_1 \psi_1} \sqrt{Q(x)} + g ^{\psi_1 \phi} \sqrt{Q(x)} cos \theta) d \psi_1 + \\ \\ +
(g^{\psi_0 \psi_1} \sqrt{Q(x)} + g^{\psi_0 \phi} \sqrt{Q(x)} cos \theta) d \psi_0 + \\ \\ +
(g^{\phi \psi_1} \sqrt{Q(x)} + g^{\phi \phi} \sqrt{Q(x)} cos \theta) d\phi,
\end{multline}

\begin{multline}
\hat{e}_5 = 2 ( g^{\psi_1 \psi_0} +  g^{\psi_1 \psi_1} ( x+ \frac{q}{x -\xi}) +  g^{\psi_1 \phi} ( x-\xi +\frac{q}{x - \xi})) d\psi_1 + \\ \\ +
2 (g^{\psi_0 \psi_0} + g^{\psi_0 \psi_1} ( x+ \frac{q}{x -\xi}) +  g^{\psi_0 \phi} (x -\xi + \frac{q}{x - \xi})) d\psi_0 + \\ \\ +
2 ( g^{\phi \psi_0} +  g^{\phi \psi_1} ( x+ \frac{q}{x - \xi}) + g^{\phi \phi} (x -\xi +\frac{q}{x - \xi})) d\phi.
\end{multline}

\noindent
And now in preparation of calculating the Ricci tensor and then the Riemann tensor, we get for $F^{ab}$ in vielbein indices:

\begin{multline}
F = \frac{2\sqrt{3} q}{x-\xi} ( \frac{1}{x-\xi} +\frac{3}{2 Q(x)} ) \hat{e}^1 \wedge \hat{e}^4 - \frac{2 \sqrt{3} q}{(x-\xi) Q(x)} \hat{e}^3 \wedge \hat{e}^4 + \\ \\ +
( - \frac{\sqrt{3} q cos \theta sin \theta }{\sqrt{Q(x)}(\xi -x)^{3/2}}) \hat{e}^3 \wedge \hat{e}^2 + \frac{2 \sqrt{3} q sin^2 \theta}{(x-\xi)^2} \hat{e}^1 \wedge \hat{e}^2.
\end{multline}

\noindent
And now we know using our metric that the only not-null components of the Ricci tensor are:

\begin{equation}
R_{aa} = -4 g_{aa} + \frac{1}{2}( g_{aa}^2 g_{cc} F^{ac} F^{ac} -\frac{1}{6} g_{aa} g_{cc} g_{dd} F^{cd} F^{cd}).
\end{equation}

\noindent
Detailing this with 4 indices (note that $ g_{aa}^{ \mu \nu} = e_{a}^{\mu} e_{a}^{\nu}$):

\begin{multline}
{R_{55}}^{ \mu \nu} = -4 g_{55}^{ \mu\nu} -\frac{1}{3}g_{55}^{\mu \nu}\{g_{11} g_{44} (F^{14})^2 + g_{11} g_{22} (F^{12})^2 + g_{33} g_{44}( F^{34})^2 + \\ \\ +
2 g_{33} g_{22} (F^{32})^2)\},
\end{multline}

\begin{multline}
{R_{44}}^{\mu \nu} = - 4 g_{44}^{\mu\nu} +\frac{1}{2}\{{(g_{44}^2)}^{\mu\nu} (g_{11} ( F^{41})^2 + g_{33}(F^{43})^2) -\\ \\ -
\frac{1}{3}g_{44}^{\mu\nu}\{g_{11} g_{44} (F^{14})^2 + g_{11} g_{22} (F^{12})^2 + \\ \\ 
+g_{33} g_{44}( F^{34})^2 + 
2 g_{33} g_{22} (F^{32})^2\}\},
\end{multline}

\begin{multline}
{R_{33}}^{\mu \nu} = - 4 g_{33}^{\mu \nu} + \frac{1}{2}\{{(g_{33}^2)}^{\mu\nu} (g_{44} ( F^{34})^2 + g_{22}(F^{32})^2) - \\ \\ -
\frac{1}{3}g_{33}^{\mu\nu}\{g_{11} g_{44} (F^{14})^2 + g_{11} g_{22} (F^{12})^2 + \\ \\ +
g_{33} g_{44}( F^{34})^2 + 
2 g_{33} g_{22} (F^{32})^2\}\},
\end{multline}

\begin{multline}
{R_{22}}^{\mu \nu} = - 4 g_{22}^{\mu \nu} + \frac{1}{2}\{{(g_{22}^2)}^{\mu\nu} (g_{33} ( F^{23})^2 + g_{11}(F^{21})^2) - \\ \\ -
\frac{1}{3}g_{22}^{\mu\nu}\{g_{11} g_{44} (F^{14})^2 + g_{11} g_{22} (F^{12})^2 + \\ \\
+ g_{33} g_{44}( F^{34})^2 + 
2 g_{33} g_{22} (F^{32})^2\}\},
\end{multline}

\begin{multline}
{R_{11}}^{\mu \nu} = - 4 g_{11}^{\mu \nu} + \frac{1}{2}\{{(g_{11}^2)}^{\mu\nu} (g_{22} ( F^{12})^2 + g_{44}(F^{14})^2) - \\ \\ -
\frac{1}{3}g_{11}^{\mu\nu}\{g_{11} g_{44} (F^{14})^2 + g_{11} g_{22} (F^{12})^2 + \\ \\+
 g_{33} g_{44}( F^{34})^2 + 
2 g_{33} g_{22} (F^{32})^2\}\}.
\end{multline}

\medskip
\noindent
And now using the formula:

\begin{equation}
{R_{\rho \sigma}}^{\mu\nu} = e_{\rho}^a e_{\sigma}^a {R_{aa}}^{\mu\nu},
\end{equation}

\noindent
we finally get to the Riemann tensor we need in the integrability conditions stated in section 5:

\begin{equation}
R_{\rho \sigma \alpha\beta} = g_{\alpha \mu} g_{\beta \nu} {R_{\rho \sigma}}^{\mu \nu}. 
\end{equation}

\medskip
\noindent
This whole context was useful to find the Riemann and the electromagnetic tensors, which appear in the integrability conditions  on the Killing spinor. In the end these integrability conditions translate in constraints on the Majorana spinors and the fact that $\chi^2 = 1$.

\medskip
\noindent
We now turn to determining the Christoffel symbols and the spin connection coefficients. The only not null Christoffel coefficients are:

\begin{multline}
\Gamma^x_{x x} = - \frac{1}{2 Q(x)^2} g^{x x}\{-8 x^3 - (1-24 \xi) x^2 + 2 \xi (1-12 \xi) x + \\ \\ +
( 8q - 2\xi + 12 \xi^2) \xi +k\},
\end{multline}

\begin{multline}
\Gamma^{\psi_0}_{\psi_1 x}  = 4 g^{\psi_0 \psi_0} ( 1 - \frac{q}{(x-\xi)^2} )+ \frac{1}{2} g^{\psi_0 \psi_1} [ -8x^3 - (1- 24\xi) x^2 + 2\xi (1 - 12 \xi) x + \\ \\ +
8 (x + \frac{q}{x-\xi})(1-\frac{q}{(x-\xi)^2})+ (8q -2\xi + 12 \xi^2)\xi +k ],
\end{multline}

\begin{multline}
\Gamma^{\psi_0}_{x \psi_1} = 4 g^{\psi_0 \psi_0} ( 1 - \frac{q}{(x-\xi)^2} )+ \frac{1}{2} g^{\psi_0 \psi_1} [ -8x^3 - (1- 24\xi) x^2 + \\ \\+ 
2\xi (1 - 12 \xi) x + 
8 (x + \frac{q}{x-\xi})(1-\frac{q}{(x-\xi)^2})+ (8q -2\xi + 12 \xi^2)\xi +k ] + \\ \\+ 
\frac{1}{2} g^{\psi_0 \phi}[ 2(-8x^3 - (1- 24\xi) x^2 +
2\xi (1 - 12 \xi) x )+ \\ \\ + 
8 (x + \frac{q}{x-\xi}) (2x -\xi +\frac{2q}{x-\xi}) + 2[(8q-2\xi + 12 \xi^2)\xi +k]],
\end{multline}

\begin{equation}
\Gamma^{\psi_1}_{x \psi_0} = 4( g^{\psi_1 \psi_1} + g^{\psi_1 \phi})(1-\frac{q}{(x-\xi)^2}),
\end{equation}

\begin{equation}
\Gamma^{\psi_1}_{\psi_0 x} = 4(g^{\psi_1 \psi_1} + g^{\psi_1 \psi_0} + g^{\psi_1 \phi})(1 - \frac{q}{(x-\xi)^2}),
\end{equation}

\begin{equation}
\Gamma^x_{\psi_1 \psi_0} = \Gamma^x_{\psi_0 \psi_1} = -4g^{xx} (1-\frac{q}{(x-\xi)^2}),
\end{equation}

\begin{multline}
\Gamma^{\psi_0}_{x \phi} = 4 g^{\psi_0 \psi_0} (1 - \frac{q}{(x-\xi)^2}) + \frac{1}{2} g^{\psi_0 \psi_1}[-2 cos\theta (-8 x^3 - (1-24 \xi) x^2 + \\ \\ +
2 \xi (1-12\xi) x ) + 8 (1- \frac{q}{(x-\xi)^2}) (2x -\xi +\frac{2q}{x-\xi}) + \\ \\ +
2 cos \theta [(8q -2\xi + 12 \xi^2) \xi +k]],
\end{multline}

\begin{multline}
\Gamma^{\psi_0}_{\phi x} = 4 g^{\psi_0 \psi_0} (1 - \frac{q}{(x-\xi)^2}) + \frac{1}{2} g^{\psi_0 \psi_1}[-2 cos\theta (-8 x^3 - (1-24 \xi) x^2 + \\ \\ +
2 \xi (1-12\xi) x ) + 8 (1- \frac{q}{(x-\xi)^2}) (2x -\xi +\frac{2q}{x-\xi}) + \\ \\+ 2 cos \theta [(8q -2\xi + 12 \xi^2) \xi +k]] - 
\frac{1}{2} g^{\psi_0 \phi} sin^2 \theta,
\end{multline}

\begin{equation}
\Gamma^x_{\psi_0 \phi} = \Gamma^x_{\phi \psi_0} = - 4 g^{xx} (1 - \frac{q}{(x-\xi)^2}),
\end{equation}

\begin{multline}
\Gamma^{\phi}_{\psi_0 x} = 4 g^{\phi\psi_0} (1 - \frac{q}{(x-\xi)^2}) +\frac{1}{2} g^{\phi \psi_1} [-2 cos\theta (-8 x^3 - (1-24 \xi) x^2 + \\ \\ +
2 \xi (1-12\xi) x ) + 8 (1- \frac{q}{(x-\xi)^2}) (2x -\xi +\frac{2q}{x-\xi}) + \\ \\+ 
2 cos \theta [(8q -2\xi + 12 \xi^2) \xi +k]] - 
\frac{1}{2}g^{\phi \phi} sin^2 \theta,
\end{multline}

\begin{equation}
\Gamma^{\phi}_{x\psi_0}  = 4 (g^{\phi \psi_1} + g^{\phi \phi}) (1 - \frac{q}{(x-\xi)^2}),
\end{equation}

\begin{multline}
\Gamma^{\psi_1}_{x \phi} = 4 g^{\psi_1 \psi_1} (1 - \frac{q}{(x-\xi)^2}) + \frac{1}{2} g^{\psi_0 \psi_1}[-2 cos\theta (-8 x^3 - (1-24 \xi) x^2 + \\ \\ +
2 \xi (1-12\xi) x ) + 8 (1- \frac{q}{(x-\xi)^2}) (2x -\xi +\frac{2q}{x-\xi}) + 2 cos \theta [(8q -2\xi + 12 \xi^2) \xi +k]] - \\ \\ -
\frac{1}{2} g^{\psi_1 \phi} sin^2 \theta,
\end{multline}

\begin{multline}
\Gamma^{\phi}_{\psi_1 x} = \Gamma^{\phi}_{ x \psi_1}= 4 g^{\phi \psi_0} (1 - \frac{q}{(x-\xi)^2}) + \frac{1}{2} g^{\phi \phi}[-2 cos\theta (-8 x^3 - (1-24 \xi) x^2 + \\ \\ +
2 \xi (1-12\xi) x ) + 8 (1- \frac{q}{(x-\xi)^2}) (2x -\xi +\frac{2q}{x-\xi}) + 2 cos \theta [(8q -2\xi + 12 \xi^2) \xi +k]] + \\ \\ +
\frac{1}{2} g^{\psi_1 \phi} [-8 x^3 - (1-24 \xi) x^2 + 2 \xi (1-12\xi) x + (8q -2\xi + 12 \xi^2)\xi +k + 8 (x+\frac{q}{x-\xi}) \\ \\
(1 - \frac{q}{(x-\xi)^2})].
\end{multline}

\medskip
\noindent
The Dirac operator is written as:

\begin{equation}
D_\mu \psi = \partial_\mu \psi + \omega_{\mu a b} \Sigma^{ab} \psi,
\end{equation}

\noindent
where the spin connections can be calculated as:

\begin{equation}
{{\omega_\mu}^a}_b = - {e^a}_\lambda {e_b}^\kappa \Gamma^{\lambda}_{\mu \kappa} -  {e^a}_\sigma \partial_\mu {e_b}^\sigma.
\end{equation}

\noindent
Note that in the end we need to calculate the following coefficients and actually only 30 are not null, out of which we need to calculate only 15, given the fact that the coefficients are antisymmetric in ab:

\begin{equation}
\omega_{\mu a b} =  g_{aa}( - {e^a}_\lambda {e_b}^\kappa \Gamma^{\lambda}_{\mu \kappa} -  {e^a}_\sigma \partial_\mu {e_b}^\sigma).
\end{equation}

\noindent
We are going to write the extended formulae (which are non-trivial) for only three of them and we are going to leave the rest in a constrained form. Here they are:

\begin{multline}
\omega_{\psi_0 4 3} =-  g^{xx} \Gamma^{\psi_1}_{\psi_0 x} = \frac{-4(\xi-x)^2}{Q(x) D^2 (\psi_0, \psi_1, x, \theta, \phi)} \{-16 Q^2 (x) cos^2 \theta +\\ \\+ 
4 Q(x) [(\xi-x) sin^2 \theta + 
32 cos \theta (x +\frac{q}{x-\xi})(x-\xi +\frac{q}{x-\xi}) - \\ \\ 
-16 (x -\xi +\frac{q}{x -\xi})^2] - 
320 (x +\frac{q}{x-\xi})^2 (x-\xi +\frac{q}{x - \xi})^2  - \\ \\ 
-48 ( x+ \frac{q}{x -\xi})^2 (\xi- x) sin^2 \theta\} 
\{ -8 Q(x) cos \theta (2x- 2\xi +\frac{2q}{x-\xi} +1) + \\ \\ +
(x+\frac{q}{x-\xi}) (\xi -x) sin^2 \theta - 4 (\xi- x) sin^2 \theta  
64 (x + \frac{q}{x-\xi})(x-\xi +\frac{q}{x-\xi})^2 - \\ \\ -
32(x+\frac{q}{x- \xi})(x-\xi +\frac{q}{x-\xi})\}(1-\frac{q}{(x-\xi)^2}),
\end{multline}

\begin{multline}
\omega_{\psi_0 5 3} =-\frac{-8(\xi-x)^2}{Q(x)^{3/2} D^2 (\psi_0, \psi_1, x, \theta, \phi)} \{-16 Q^2 (x) cos^2 \theta +\\ \\+ 
4 Q(x) [(\xi-x) sin^2 \theta + 
32 cos \theta (x +\frac{q}{x-\xi})(x-\xi +\frac{q}{x-\xi}) - 16 (x -\xi +\frac{q}{x -\xi})^2] - \\ \\ -
320 (x +\frac{q}{x-\xi})^2 (x-\xi +\frac{q}{x - \xi})^2  - 48 ( x+ \frac{q}{x -\xi})^2 (\xi- x) sin^2 \theta\} \\ \\
\{ -8 Q(x) cos \theta (2x- 2\xi +\frac{2q}{x-\xi} +1) + (x+\frac{q}{x-\xi}) (\xi -x) sin^2 \theta - \\ \\-
4 (\xi- x) sin^2 \theta  -64 (x + \frac{q}{x-\xi})(x-\xi +\frac{q}{x-\xi})^2 - \\ \\ -
32(x+\frac{q}{x- \xi})(x-\xi +\frac{q}{x-\xi})\}(1-\frac{q}{(x-\xi)^2})(x+\frac{q}{x-\xi})\},
\end{multline}

\begin{multline}
\omega_{\psi_0 3 2} = \frac{16(\xi-x)^{\frac{7}{2}} sin \theta}{D^3 (\psi_0, \psi_1, x, \theta, \phi) Q^2(x)}\{-16 Q^2 (x) cos^2 \theta +
4 Q(x) [(\xi-x) sin^2 \theta + \\ \\
32 cos \theta (x +\frac{q}{x-\xi})(x-\xi +\frac{q}{x-\xi}) - 16 (x -\xi +\frac{q}{x -\xi})^2] - \\ \\ -
320 (x +\frac{q}{x-\xi})^2 (x-\xi +\frac{q}{x - \xi})^2  - 48 ( x+ \frac{q}{x -\xi})^2 (\xi- x) sin^2 \theta\} \\ \\
\{ -8 Q(x) cos \theta (2x- 2\xi +\frac{2q}{x-\xi} +1) + (x+\frac{q}{x-\xi}) (\xi -x) sin^2 \theta - \\ \\-
4 (\xi- x) sin^2 \theta  -64 (x + \frac{q}{x-\xi})(x-\xi +\frac{q}{x-\xi})^2 - \\ \\ -
32(x+\frac{q}{x- \xi})(x-\xi +\frac{q}{x-\xi})\} (1-\frac{q}{x-\xi})^2.
\end{multline}

\medskip
\noindent
And now the rest of the spin connections in constrained form:

\begin{equation}
\omega_{\psi_1 5 3} = - \frac{2 g^{xx}}{\sqrt{Q(x)}} \Gamma^{\psi_0}_{\psi_1 x},
\end{equation}

\begin{equation}
\omega_{\psi_1 3 4}=  - \frac{g^{xx}}{Q(x)}(g^{\psi_0 \psi_1} + g^{\psi_0 \phi} cos\theta) \Gamma^{x}_{\psi_1 \psi_0},
\end{equation}

\begin{equation}
\omega_{\psi_1 3 2} = - \frac{\sqrt{\xi -x} sin \theta}{Q(x)^{\frac{3}{2}}} g^{xx} g^{\psi_0 \phi} \Gamma^{x}_{\psi_1 \psi_0},
\end{equation}

\begin{equation}
\omega_{\phi 5 3} = - \frac{2 g^{xx}}{\sqrt{Q(x)}} \Gamma^{\psi_0}_{\phi x},
\end{equation}

\begin{equation}
\omega_{\phi 3 4} = - \frac{g^{xx}}{Q(x)}{Q(x)}(g^{\psi_0 \psi_1} + g^{\psi_0 \phi} cos\theta) \Gamma^{x}_{\phi \psi_0},
\end{equation}

\begin{equation}
\omega_{\phi 3 2} = - \frac{\sqrt{\xi -x} sin \theta}{Q(x)^{\frac{3}{2}}} g^{xx} g^{\psi_0 \phi} \Gamma^{x}_{\phi \psi_0},
\end{equation}

\begin{multline}
\omega_{x 5 4} = 2 \sqrt{Q(x)} (x+\frac{q}{x-\xi}) (g^{\psi_0 \psi_1} + g^{\psi_0 \phi} cos \theta) \Gamma^{\psi_1}_{x \psi_0} + \\ \\ + \{ 16 (g^{\psi_0 \psi_0} + g^{\psi_0 \psi_1} ( x+ \frac{q}{x -\xi}) +  g^{\psi_0 \phi} (x -\xi + \frac{q}{x - \xi}))^2 + \\ \\ +
4  ( g^{\phi \psi_0} +  g^{\phi \psi_1} ( x+ \frac{q}{x - \xi}) + g^{\phi \phi} (x -\xi +\frac{q}{x - \xi}))^2  (\xi - x) sin^2 \theta + \\ \\+
4 ( g^{\psi_1 \psi_0} +  g^{\psi_1 \psi_1} ( x+ \frac{q}{x -\xi}) +  g^{\psi_1 \phi} ( x-\xi +\frac{q}{x - \xi}))^2[Q(x) + 4( x+ \frac{q}{x-\xi})^2] + \\ \\+
32 ( g^{\psi_1 \psi_0} +  g^{\psi_1 \psi_1} ( x+ \frac{q}{x -\xi}) +  g^{\psi_1 \phi} ( x-\xi +\frac{q}{x - \xi})) \\ \\
(g^{\psi_0 \psi_0} + g^{\psi_0 \psi_1} ( x+ \frac{q}{x -\xi}) +  g^{\psi_0 \phi} (x -\xi + \frac{q}{x - \xi})) (x+ \frac{q}{x-\xi}) + \\ \\+
32 (g^{\psi_0 \psi_0} + g^{\psi_0 \psi_1} ( x+ \frac{q}{x -\xi}) +  g^{\psi_0 \phi} (x -\xi + \frac{q}{x - \xi})) \\ \\ 
( g^{\phi \psi_0} +  g^{\phi \psi_1} ( x+ \frac{q}{x - \xi}) + g^{\phi \phi} (x -\xi +\frac{q}{x - \xi})) (x- \xi +\frac{q}{x-\xi})\} \\ \\
\{( g^{\psi_1 \psi_1} \sqrt{Q(x)} + g ^{\psi_1 \phi} \sqrt{Q(x)} cos \theta) + (g^{\phi \psi_1} \sqrt{Q(x)} + \\ \\+
g^{\phi \phi} \sqrt{Q(x)} cos \theta) cos \theta\} ((1-\frac{q}{(x- \xi)^2}).
\end{multline}

\medskip
\noindent
We write the following two in constrained form noting that the term in $g_{55}$ writes the same way in both the two related terms:

\begin{multline}
\omega_{x 5 2} = \sqrt{\xi - x} sin \theta \{ -2 (x+\frac{q}{x-\xi}) g^{\psi_0 \phi} \Gamma^{\psi_1}_{x \psi_0} + \\ \\ +
e^\mu_5 e^\nu_5 g_{\mu\nu} ( g^{\phi \phi} + g^{\psi_1 \phi}) (1 - \frac{q}{(x-\xi)^2})\},
\end{multline} 

\begin{multline}
\omega_{x 4 2} =\sqrt{\xi - x} sin \theta \{\sqrt{Q(x)} g^{\psi_0 \phi} \Gamma^{\psi_1}_{x \psi_0} + \\ \\ +
\frac{\partial_x Q(x)}{2 \sqrt{Q(x)}} e^\mu_4 e^\nu_4 g_{\mu\nu}(g^{\psi_1 \phi} + g^{\phi \phi} cos \theta) \}.
\end{multline}

\medskip
\noindent
The last three spin connection coefficients are:

\begin{equation}
\omega_{\theta 2 4} =\sqrt{(\xi-x)Q(x)} cos \theta e^{\mu}_2 e^{\nu}_2 g_{\mu \nu} (g^{\phi \psi_1} + g^{\phi \phi} cos \theta),
\end{equation}

\begin{multline}
\omega_{\theta 2 5} = 2 \sqrt{\xi -x} cos \theta [g^{\phi \psi_0} + g^{\phi \psi_1} (x + \frac{q}{x-\xi}) + \\ \\ +
g^{\phi \phi} (x - \xi + \frac{q}{x-\xi})] e^\mu_2 e^\nu_2 g_{\mu \nu},
\end{multline}

\begin{multline}
\omega_{\theta 4 5} = -\sqrt{Q(x)} sin \theta [g^{\phi \psi_0} + g^{\phi \psi_1} (x + \frac{q}{x-\xi}) +  \\ \\ +
g^{\phi \phi} (x - \xi + \frac{q}{x-\xi})] e^\mu_4 e^\nu_4 g_{\mu \nu}.
\end{multline}


\end{document}